# Time-Aware Music Recommender Systems: Modeling the Evolution of Implicit User Preferences and User Listening Habits in A Collaborative Filtering Approach

**Diego Sánchez-Moreno** [1], **Yong Zheng** [2] and **María N. Moreno-García** [1,*]

1. Department of Computer Science and Automation, University of Salamanca, 37008 Salamanca, Spain; sanchezhh@gmail.com
2. Department of Information Technology and Management, Illinois Institute of Technology, Chicago, IL 60616, USA; yzheng66@iit.edu
* Correspondence: mmg@usal.es

**Abstract:** Online streaming services have become the most popular way of listening to music. The majority of these services are endowed with recommendation mechanisms that help users to discover songs and artists that may interest them from the vast amount of music available. However, many are not reliable as they may not take into account contextual aspects or the ever-evolving user behavior. Therefore, it is necessary to develop systems that consider these aspects. In the field of music, time is one of the most important factors influencing user preferences and managing its effects, and is the motivation behind the work presented in this paper. Here, the temporal information regarding when songs are played is examined. The purpose is to model both the evolution of user preferences in the form of evolving implicit ratings and user listening behavior. In the collaborative filtering method proposed in this work, daily listening habits are captured in order to characterize users and provide them with more reliable recommendations. The results of the validation prove that this approach outperforms other methods in generating both context-aware and context-free recommendations.

**Keywords:** time-aware music recommender systems; TARS; CARS; implicit ratings; collaborative filtering

## 1. Introduction

The arrival of streaming services has changed the way users listen to music. Currently, music listeners do not need to store the songs they want to listen to but can access and play an almost unlimited number of songs online. This fact has given rise to the need for endowing these systems with personalized recommendation mechanisms that make it easier for users to find songs that suit their tastes. Most streaming music platforms have increasingly advanced personalization tools to offer music to users and even to show them ads. This is a key aspect of the platform's positioning against their competitors and increases their revenues. To get an idea of the economic power of these platforms, Spotify, one of the most popular platforms, reached a revenue of 2 billion USD in the first quarter of 2020, 161 million USD of which was due to ads [1].

For several years, research on recommender systems has been very intense, and more recently the focus has been centered on context-aware systems, which aim to adapt recommendations according to the circumstances surrounding the user. To do so, different contextual factors, such as the location, time, weather, companions, and emotions, among others, are added as additional dimensions to the recommendation models. Moreover, time is an essential factor since it not only has an effect on the user context, but also influences the drift of user preferences.

In the music domain, dealing with the effect of time on recommendations is different from other fields, especially in the context of streaming services. Therefore, approaches used in some domains may be difficult to be reused or extended to this area. This is mainly due to the differences in the way music is consumed compared to other products. A music item is usually played many times by the same user, while a book is generally read only once. Another difference is the number of items that can be consumed in a given period, which can be high for music and lower for other items such as books. Likewise, the evolution of preferences may be different depending on the application domain since in some domains, interests evolve quickly, while in others these changes occur more slowly. Music, on the other hand, is found somewhere in the middle, where changes occur at a more even pace. In addition, when considering time as a contextual variable, its effects on the field of music are quite different from those generated in other areas. For instance, the time of day can have little influence when choosing a book but can have a significant influence when choosing a song.

The previously described issues may give rise to additional problems when explicit ratings are not available, and it becomes necessary to resort to implicit feedback. The common methods for obtaining implicit ratings, such as using the number of records purchased or the time spent examining items, are not suitable for the music domain. The reason for this is that music items are not individually purchased on streaming platforms, and that the same item can be listened to many times by the same user. This forces music preference evolution, managed from the perspective of implicit feedback, to be different from other areas. Quite often, the number of times a song is played is used to derive implicit ratings about a particular musical item. However, computing ratings over time, as a means to manage the drift in preference, requires establishing time windows to distribute timestamped plays and compute the evolving ratings for each interval. In other domains, the procedure is more straightforward, since only an observation per user and item is needed to apply a decay function or to discard old ratings. The observation can correspond to an action, such as a purchase. Nevertheless, timestamped ratings of music items require several observations in the form of plays for each time window. The same problem arises in context-aware recommender systems when setting time contexts.

The drawbacks inherent to music recommender systems with respect to time dynamics are addressed in this work. The proposed method manages the evolution of user preferences in a continuous way by using implicit ratings and avoiding time windows. Our proposal also exploits user listening behavior to improve recommendations. The validation of this approach has been performed through a case study, where a music dataset is used to recommend songs, both when the context is not considered and when the time of day context is taken into account. The research presented in this paper is an extension of previous work in which contextual aspects were not addressed [2].

The main contributions of this work are the following:

- A method to induce music implicit ratings that addresses the evolution of user preferences.
- A way of incorporating time-dependent user listening behavior in user-based collaborative filtering models.
- An approach for providing context-aware recommendations that combine the evolution of user preferences, user listening behavior and the time of day as a contextual factor.

To conclude this section, we refer to other important aspects of streaming services that are worthy of consideration, although an in-depth analysis is beyond the scope of this work. The first is related to the security of cloud services used on-demand in the context of smart cities. Different security issues must be implemented for maintaining a trusted environment [3]. Privacy is the second aspect to be considered, which is also a crucial issue of trusted cloud services. The growing concern about preserving the privacy of users demands the implementation of mechanisms to prevent access to the identity of users and their personal information [4]. These two aspects are especially important in applications such as recommender systems that handle personal data and user behavior information, even though these are treated anonymously.

The rest of the paper is organized as follows. Section 2 is dedicated to introducing background information and describing related work. In Section 3, the proposed approach for addressing time-awareness in the area of music is presented. The experimental study and its results are reported in Section 4. Finally, the conclusions are given in Section 5.

## 2. Background and Related Work

### 2.1. Collaborative Filtering

There are currently many proposals for recommendation methods, most of which are based on collaborative filtering (CF). This approach consists of making predictions about the rating that a user would assign to an item from the ratings of other users who have similar preferences. It requires either implicit or explicit knowledge of user personal preferences for items in the form of ratings.

Collaborative filtering methods can be classified into memory-based and model-based methods. Memory-based methods fall into the group of lazy learning methods, which do not require a training phase since the calculations needed to make recommendations are specific to the instance of recommendation at each particular time by using all the ratings available in the system. On the contrary, in model-based methods, recommendations are obtained from a model that has been induced prior to the recommendation time through a training phase. In the first group, the nearest neighbor algorithms are commonly used to compute the similarities between users or between items, in which case we can differentiate between user-based and item-based methods [5]. The last approach can be also used as a model-based schema since rating-based similarities between items can be computed before the recommendation time [6]. In this scenario, the active user is recommended items similar to those he/she previously rated positively. Item, rather than user, similarity models are used because the former are more stable over time than the latter. It is expected that new ratings given to items in large databases do not significantly change the similarity between them, while the change in user preferences is more noticeable. User-based collaborative filtering methods provide the active user with recommendations based on the ratings of his/her neighbors. Two users are neighbors if they have some items in common that they have rated with similar scores. Several metrics can be applied to computer user similarities, although the Pearson correlation coefficient and cosine similarity are the most frequently used due to good results. The ratings given to products by the nearest neighbors and similarity coefficients are used to predict the rating that a given user would assign to an item that he or she has not yet rated.

The process is formally described as follows. Let $U = \{u_1, u_2, \ldots, u_m\}$ be a set of $m$ users and $I = (i_1, i_2, \ldots, i_n)$ a set of n items, where each user $u_i$ has a list of ratings that he has given to a set of items $I_{ui}$, where $I_{ui} \subseteq I$. The ratings are stored in a $m \times n$ matrix called the rating matrix, where each element is the rating that a user $u_i$ gives to an item $i_j$. Given that users have not rated all the items, this matrix has some empty elements, in such a way that the fewer the number of rated items the sparser the matrix is. This can be a problem that is typical of collaborative filtering (CF) methods. If the number of ratings is high enough, recommendations can be made to the active user $u_a$ by calculating the similarities with the other users in the system who have rated the same products.

The Pearson correlation coefficient, $\omega(u_a, u_i)$, can be used to compute the similarity between the active user $u_a$ and another user $u_i$. This coefficient evaluates the linear relationship between two variables and is obtained from its covariance according to Equation (1) [7].

$$\omega(u_a, u_i) = \frac{\sum_j (r_{aj} - \bar{r}_a)(r_{ij} - \bar{r}_i)}{\sqrt{\sum_j (r_{aj} - \bar{r}_a)^2 (r_{ij} - \bar{r}_i)^2}} \tag{1}$$

where $r_{aj}$ and $r_{ij}$ are the ratings that $u_a$ and $u_i$ give to item $i_j$ respectively, and $\bar{r}_a$ and $\bar{r}_i$ are the average ratings of $u_a$ and $u_i$, respectively. The Pearson coefficient can represent inverse and direct correlation, with values in the interval [−1, 1], where the value 0 corresponds to the absence of a correlation.

Another commonly used similarity metric is cosine, which is given by the dot product of the vectors representing the preferences of two given users $u_a$ and $u_i$ in the Euclidean space. The cosine similarity between those users, $cosSim(u_a, u_i) = cos(R_{u_a}, R_{u_i})$, is computed according to Equation

(2) [8], where $R_{u_a}$ and $R_{u_i}$ are the vectors containing the ratings given to items by users $u_a$ and $u_i$, respectively.

$$cos\left(R_{u_a}, R_{u_i}\right) = \frac{R_{u_a} \cdot R_{u_i}}{\|R_{u_a}\| \|R_{u_i}\|} = \frac{\sum_{j=1}^{n} r_{aj} r_{ij}}{\sqrt{\sum_{j=1}^{n} r_{aj}^2} \sqrt{\sum_{j=1}^{n} r_{ij}^2}} \quad (2)$$

The ratings of the *k* of most similar users are used to estimate the preferences of the active users about items that they have not rated. The rating $pr_{aj}$ estimated for the active user $u_a$ and an item $i_j$ is computed using the following equation [8]:

$$pr_{aj} = \bar{r}_a + \frac{\sum_{i=1}^{k} sim(u_a, u_i)(r_{ij} - \bar{r}_i)}{\sum_{i=1}^{k} |sim(u_a, u_i)|} \quad (3)$$

where $sim(u_a, u_i)$ is the similarity between the active user and his neighbor $u_i$, which could be obtained by using the Pearson coefficient, cosine or any other similarity metric.

There are other ways of computing ratings from similarities, especially when they are used for top-N recommendations. These are given in the form of a ranked list containing N items. In this context, the normalizing denominator can be eliminated, since there is no need for the ratings to be adjusted to a range. Rating averages can also be replaced by bias for the user–item pairs [9] or eliminated [10]. Other memory-based CF methods involve changes in the prediction function, such as rating normalizations using Z-score [11]. However, most of the proposals are hybrid approaches [12] that use the ratings and information from users or items. This additional information can be item content data (e.g., the genre of a movie) [13] or user data such as demographic attributes [14], user social data [15,16], and other types of user information [17] which is usually used in combination with ratings to compute similarity.

In the category of model-based methods, there are many proposals in the literature. Some of these focus on replacing the use of the lazy k-nearest neighbor technique with other machine learning algorithms [18], such as associative classification [19], rule-based methods [20,21], Bayes classifiers [22], support vector machines [23], and neural networks [24].

Latent factor models [25] are more widely used than approaches based on machine learning. This group includes matrix factorization [26], which is a procedure for dimensionality reductions that generate latent factors for each user and each item. It consists of decomposing the rating matrix $R \in \mathbb{R}^{n \times m}$ with many empty elements into two less scarce matrices, a user matrix $U \in \mathbb{R}^{m \times d}$ and an item matrix $V \in \mathbb{R}^{n \times d}$, so that the product $UV^T$ is an approximation of R. The latent factors of $U$ and $V$ will be those that minimize the function $\sum_{i,j}(R_{i,j} - \langle U_i, V_j \rangle)^2$.

A popular factorization technique in the area of recommender systems is singular value decomposition (SVD) [27]. SVD is particularly suitable for dealing with the sparsity problem that occurs when the number of ratings is low. In this scenario, SVD yields more reliable recommendations than standard CF algorithms [6], although it may suffer from overfitting problems. Moreover, it has a high computational cost in large-scale systems; thus, less expensive SVD-based approaches, such as incremental SVD have been proposed [28]. Other SVD-based techniques for dimensionality reductions, such as latent semantic indexing (LSI) [29], are also indicated for the same purpose. In recent years, different recommendation methods based on matrix factorization have been developed [30,31], including SVD++ [32], which incorporates implicit feedback. Other variants of these methods that consider time aspects are described in the following section.

## 2.2. Time-Aware Recommender Systems

Time can influence user preferences in different ways and has been widely studied in the literature. Systems considering these temporal effects are commonly known as time-aware recommender systems (TARS). They have been classified according to the type of influence being considered, although the categories established by different authors are not the same. Most of these systems differentiate between the evolution of preferences over time and their change as a consequence of the temporal context (time of day, workday or weekend, season, etc.).

Within CF approaches, models to address time effects can be classified in three groups [18]: recency-based models, periodic context-based models and models that explicitly use time as an independent variable. Recency-based models give more importance to recent ratings than to older ones in CF. Depending on the way the ratings are managed, this category can be subdivided in window-based and decay-based models. Periodic context-based models are context-aware approaches where the only contextual dimension is time, which is split into intervals corresponding to diverse temporal contexts (winter/spring/summer/autumn, weekend/workday, morning/evening, etc.). The third group includes more advanced methods for modeling user and item trends.

In [33], three categories of time-aware recommender systems are defined: continuous time-aware, categorical time-aware, and time-adaptive methods. The first category is analogous to recency-based models that capture the evolution of preferences. It is usually done by means of functions that increase the weight of the ratings as they approach the target recommendation time. Categorical time-aware systems implement temporal context-aware methods. In time-adaptive systems, the more recent the ratings, the more important they are, which is similar to continuous time-aware methods, but without considering any target recommendation time.

Recently, temporal dynamics have been addressed in some studies with regard to the repeat of consumption. These approaches are suitable for application domains where the same items are consumed more than once (purchasing a product, listening to a song, visiting a place, etc.). While some proposals are only focused on predicting when the users will repeat the consumption of items [34], others provide recommendations of items when a repetition is predicted [35]. In addition, the short-term and long-term time fluctuations of repeat consumption have also been modeled [30]. Although these approaches may be useful in the music field, the objectives are very different from those of our proposal. Our approach falls within the scope of CF, where new products that the user has not previously consumed are recommended. In [36], CF is also considered as a recommendation, but the temporal dynamics are restricted to consumption repetition. Preference evolution, which is one of the main concerns of our work, is not considered.

2.2.1. Evolution of User Preferences

Proposals for addressing preference evolution commonly require explicit ratings and information about the time when they were obtained from users. Timestamped ratings are necessary to adjust their weight [37], using either decay functions or time windows.

In the group of recency-based models, decay-based models usually involve exponential functions of time. Let us consider a target time $t_f$ representing a future time when recommendations will be given, and $t_{u,j}$ the time when the user $u$ rates the item $j$. The importance of every rating at the recommendation time is given by a weight $w_{u,j}$ computed from a decay function $f(t)$ that decreases the weight of the ratings as their timestamp moves away from the target time [18]. Ding and Li [38] proposed the following decay function to reduce monotonically rating importance with time in the range (0, 1).

$$w_{u,j}(t_f) = f(t) = e^{-\lambda (t)} \qquad (4)$$

where $\lambda$ is the decay rate and $t = t_f - t_{u,j}$.

The decay rate $\lambda$ is defined as the inverse of the half-life parameter $T_0$ (Equations (5) and (6)).

$$\lambda = \frac{1}{T_0} \qquad (5)$$

The parameter $T_0$ is defined by Equation (6), which indicates a reduction in the weight by half in $T_0$ days after the initial time $t = 0$.

$$F(T_0) = \frac{1}{2} f(0) \qquad (6)$$

In order to predict the rating $pr_{aj}$ for the active user $u_a$ and the item $j$ at the target time $t_f$ in user-based CF, Equation (3) is modified by introducing the weights $w_{u_i,j}(t_f)$ of the ratings given by the $k$ neighbors $u_i$ of the active user who have rated the item $j$:

$$pr_{aj} = \bar{r}_a + \frac{\sum_{i=1}^{k} w_{u,j}(t_f) \cdot sim(u_a, u_i)(r_{ij} - \bar{r}_i)}{\sum_{i=1}^{k} |sim(u_a, u_i)|} \tag{7}$$

This approach can also be used in item-based CF by incorporating the weight in a similar way to the corresponding formula.

In window-based models, corresponding to the second subdivision of the category of recency-based models, a time interval preceding the target time is established. Then, ratings with a timestamp outside the time window are discarded [39,40]. The size of the window depends on the application domain.

An approach combining time windows and decay-based models is presented in [41], where time is partitioned in intervals and the users' interest distribution in these intervals is analyzed in order to model its evolution. As a result, the following decay function, Equation (8), for an elapsed time *t* was proposed:

$$f(t) = 0.5 + 0.5 \times e^{-\lambda t} \tag{8}$$

The decay rate $\lambda$ is obtained from Equation (9), where $N_c$ is the number of preferences of catalog c, and $n$ is the total number of ratings. Moreover, $\alpha$ is a constant whose value lies between 0 and 1, and $\sum_{i=1}^{k} d(i)$ is the number of rated items during one period.

$$\lambda = \alpha \sum_{i=1}^{k} d(i) \times N_c / n \tag{9}$$

In [42], the window and decay-based approaches are also used together in a bipartite network-based recommendation method. They define the short- and long-term windows from the timestamp information associated to explicit ratings, then an exponential attenuation function is applied to the ratings in order to decrease their importance at a different speed in each long-term window. The method is validated with two MovieLens datasets of different sizes, showing a better performance than other network-based algorithms.

Using time as an independent variable allows us to model preference trends with respect to both users and items. Although time-series methods can be applied, the most extended approaches are temporal factor models. They have the advantage over the previously described techniques in that they allow noise and temporary trends to be differentiated from long-term trends, but with more complexity [18].

The methods described above can be considered as the background approaches for dealing with the evolution of preferences; however, other proposals have been reported in the literature. Some methods can be placed in some of the established groups of techniques, but others are hybrid approaches that include characteristics of more than one category. Some representative papers are analyzed below.

Hermann presented in [43] a method for recommending lecture materials, in which the time similarity between items is computed considering their download dates and the recommendation time $t$. The similarity metric for two items $i$ and $j$ is defined in Equation (10), in which the second addend in the denominator is introduced to penalize the less recent ratings.

$$S_{i,j}(t) = \frac{1}{|t_i - t_j| + |\min(t_i, t_j) - t|} \tag{10}$$

This proposal is a recency approach designed for item-based CF. Similar to the problem at hand in this paper, no explicit ratings are needed, since the downloading of materials is used as implicit user feedback. However, the extension of this approach to the field of music would be complicated due to the fact that a certain material is usually downloaded only once by a given user, while a song is usually played multiple times by the same person.

A recency-based method is also proposed in [44], where the interest drift over time is captured by means of a linear function. This function needs to be computed for the different time bins into which the full time period under consideration is divided. This proposal has been designed for the specific scenario of one-class collaborative filtering, which takes place when only positive (e.g., like) or unknown user–item feedback is available. Often, missing values are related to not-liked items, but this is not always true. Aiming at avoiding this problem, a variant of the LDA (latent dirichlet allocation) model is used for capturing latent topics from the user–item rating matrix and identifying users' latent interests. Although LDA is usually used in the text processing area, in the context of this work, documents are replaced by users and words by items. This method is only applicable when timestamped explicit feedback is available. In addition, time intervals must be created, unlike our approach, in which it is not necessary.

In [45], the preference changes over time are addressed by identifying overlapping communities among users from a time-weighted user similarity matrix. To create this matrix, timestamped ratings or transactions are used to split the dataset into several subsets for different time periods. When computing the similarity between users, these time intervals are taken into account since it is expected that the similarity is greater when they have common items rated in the same period. The main objective of grouping similar users into overlapping clusters is to overcome sparsity and cold-start problems. Recommendations are based on a model of association rules that are mined from each community. The data involved in these rules are binary ratings relating users and items, that is, the information about items that have been rated or purchased by the corresponding users. Subsets corresponding to different time partitions are needed again to introduce time effects, and a time-based confidence measure is defined to determine which are the best rules for making recommendations. The proposal is validated with two movie datasets. Although transactions can be used as implicit feedback in this work, only binary values are managed (whether an item has been purchased/rated or not) for both implicit and explicit ratings. Our objective is to work with multivalued ratings since they provide more information. Regarding temporal factor models for capturing trends, one of the most known methods is time-SVD++ [46], which introduces a modification in the original SVD++ matrix factorization method. Time effects are included in some terms already defined in SVD++ (user biases, item biases and user factors), but are handled as time-dependent variables. In [47], a Bayesian probabilistic tensor factorization model is proposed, where a feature vector including the time variable is associated with each time step. Although SVD++ can be used with implicit ratings, time-SVD++ requires explicit ratings.

Matrix factorization is also used in a proposal for points of interest (POI) recommendations that consider time windows [48]. It incorporates recency issues and comparative choices for learning user preferences since it is assumed that users give ratings to POIs by comparison to other POIs recently visited by them. The POIs in each time window are sorted according to their ratings in a partial order and used to induce a choice model including latent factors, which are learned by means of a stochastic gradient descent algorithm. This is a model that requires timestamped ratings.

A latent factor model based on matrix factorization is proposed in [49]. The proposal is an improvement on a previous method by the same authors that captures the explicit influence of the social relationships on each one of the aspects of preference. The new method incorporates the temporal dynamics of preference aspects by means of time-dependent functions. The results of the testing on three datasets about movies showed a significant increase in performance over the prior method. A similar work is presented in [50], where trust relationships between users and time information are included in an SVD-based method. The experiments were conducted on a dataset containing the ratings and reviews of movies, as well as the ratings on the reviews from which trust relationships are extracted. Time dynamics are incorporated into user and item bias but the whole time period is split into three stages (the fluctuating, approaching and stable stage) in order to adapt the rating evolution to the patterns found in a preliminary study. These methods also use timestamped explicit ratings.

Moreover, the number of studies analyzing the evolving preferences over time in music recommender systems is low. In this regard, in [2], a recency-based model for recommending songs is proposed to capture the changes in user preferences over time when explicit ratings are not available. In the category of temporal factor models, a proposal is presented in [51], where an item taxonomy is used to modify the basic bias approach. The time information linked to the ratings given to songs is used to identify sessions that are used to model the temporal dynamics of users and items. User behavior regarding the ratings assigned to songs is derived from the position of the songs during the session. The item popularity drift over time is modeled using a bias model. Apart from using explicit ratings, the main drawback of this method is the need for classifying items in a taxonomy.

Chen et al. [52] proposed a music recommendation method where the recency effects on user interest are considered. The authors introduce the concept of an interest forgetting curve (IFC) to model such time effects. They use the transition probabilities between songs in past playlists listened to by users, in order to recommend the top-k most probable songs to play next during a given time within the user's current playlist. Two ways of computing these probabilities are proposed: sequence-based transition probabilities and IFC-based transition probabilities. In the latter, a modification of the former is introduced to consider the decreasing interest over time. They suggest an exponential decreasing similar to our approach, but it is not used in the context of collaborative filtering. Another difference is the fact that recommendations are made when the user is listening to a playlist.

In [53], a CF method is proposed, but instead of using explicit ratings, the similarity between users is computed from the tags they assign to items. The proposal is validated with datasets from several domains, music being one of them. The authors report that their method is based on the assumption that two users have similar tastes if they give the same tags to the same items. They also consider that the closer in time the tag assignments are, the greater the degree of similarity between users. In this work, several decay functions are examined, but in the experimental study, the similarity is computed from a binary implicit rating (the user bookmarked the item or not) without considering tag content. The application of this method can be affected by the sparsity drawback, since only a small portion of users gives tags to items. In addition, not all systems have tagging functions.

Due to their successful application in many domains, the growing popularity of deep learning algorithms has meant that these techniques are now also being used in the area of recommender systems, especially when the process involves image processing. This is the case with the work of Yang and Zhang [54], where visual features are integrated with other kinds of attributes in a recommendation method especially indicated for large-scale systems and application areas, such as fashion, where the visual characteristics of the items are critical. The method addresses the quick changes in preferences in this field through a complex procedure that involves a time-dependent function to adapt to the dynamic changes of visual features, which are extracted from images by means of a convolutional neural network (CNN).

Periodic context-based models can be considered as context-aware recommender schemas where time is the context. Thus, recommendations are given for a specific context formed by one or more dimensions corresponding to contextual variables (time, location, weather, companion, mood, etc.). The next section is devoted to these approaches.

Dealing with time dynamics is a problem that is present in many application domains and therefore multiple methods are proposed to address it. In addition to the approaches described before, evolutionary and nature-inspired algorithms [55] [56] have also been used for this purpose in fields such as electrical energy distribution, renewable energy sources or biomedical signaling and image processing, among others. These methods and other advanced algorithms [57] are generally aimed at solving more complex problems than the one presented in this study. Thus, they are beyond the scope of this paper.

2.2.2. Temporal Context

In context-aware systems, contextual dimensions are added to the item and user dimensions involved in traditional CF models. Then, traditional 2D models are replaced by 3D models with the aim to adapt recommendations to the specific context of the user, since preferences of users for a given item can change depending on contextual variables. In the case of time-aware recommender systems, time is the contextual dimension, which is modeled as a discrete variable representing the context (time of day, day of the week, season, etc.).

Three strategies can be applied in order to make recommendations for specific contexts: contextual pre-filtering, contextual post-filtering and contextual modeling. Pre-filtering methods apply 2D CF, but only use the ratings associated with the target context. For instance, in order to recommend movies to watch at the weekend, only the ratings of movies watched on weekend are used. Contextual post-filtering methods initially ignore contextual information and recommendations are obtained using traditional 2D approaches. Then, several procedures can be used to select some items to be recommended in the target context or to adjust the ranking of these items, making use of contextual information. In contextual modeling, a 3D model incorporating contextual information is used to make recommendations. One way to do this is to compute the similarity between users or items, not only taking into account the ratings, but also contextual factors such as time.

One of the main drawbacks of pre-filtering methods is the sparsity, which is caused by the fact that ratings irrelevant to a given target context are removed. Additionally, post-filtering methods usually require the use of content information to provide contextual recommendations. For example, if a movie genre or the movies of a specific actor/director are often watched on the weekends, this content information (genre, actor, etc.) is used to adjust recommendations to the context of the user [18].

There are many studies where time is used as a contextual variable in many domains. The time of the year is considered in a comparative study of pre- and post-filtering approaches [58] conducted using a dataset of electronic products that are sold through an e-commerce system. A user-based CF method is applied to compare the results of both techniques. Categories of products are used instead of individual items, and implicit ratings are computed from the frequencies of purchases. The authors conclude that, although the results depend on the specific method of each filtering type, good post-filtering approaches should be better than pre-filtering. Time-aware recommender systems have also been developed in the field of services to help users discover the most suitable ones [59]. Recommendations are based on historical service quality (QoS) data that are usually distributed and owned by different companies, who are unable to make this information public so as to protect user privacy. In [59], the locality-sensitive hashing (LSH) technique has been modified by incorporating slots referring to the time when users invoke the services. LSH was introduced into service recommendations as a means to integrate multi-source data while preserving privacy.

Time combined with additional contextual factors has been widely studied. For example, the place, time and company are the contextual variables involved in the film recommendation system presented in [60]. In this work, a multidimensional approach is proposed to jointly manage information regarding users, items and contextual information. The ratings are estimated using a reduction-based approach that applies the classical CF technique on contextual segments. In recent years, the improvement of recommendations through the exploitation of contextual information has been the focus of intensive research [61]. The possibility of obtaining this information through mobile devices has led to the development of context-aware recommendation systems in many application domains. Tourism is one of these domains, given that data regarding factors like location, time and weather, obtained through these devices, are essential for recommending activities or tourist points of interest [62,63].

In most of the work involving this category, time is only studied from the point of view of the context, without considering time dynamics. Some proposals [64,65] consider three temporal aspects, the periodicity, consecutiveness, and non-uniformness, for making point of interest (POI) recommendations in location-based social networks (LBSNs). In [64], an aggregated temporal tensor factorization (ATTF) model is proposed to capture those temporal features. The recommendations of POIs at a specific time are based on user patterns involving the temporal features. The POI check-ins made by users on the social network are used as implicit feedback. The same temporal aspects are studied in [65], where the concept of a temporal subset property (TSP) is introduced to designate the time dimensions comprised for granular slots, where some of them being subsets of others. In this work, the multi-aspect time-related influence (MATI) probabilistic model is proposed for location recommendations. It is a multivariate model that simultaneously considers multiple latent temporal parameters. These methods are highly focused on aspects related to points of interest (restaurants, bars, etc.), as well as on the behavioral patterns of users when visiting these places. Therefore, it would be difficult to extend the use of these models to other application domains. Furthermore, although some aspects of temporal dynamics are considered, the evolution of user preferences is not one of them.

In the field of music, contextual aspects such as the time of day, season, weather or mood, have a particular importance with respect to the type of music listened to by the user. Therefore, these contextual conditions should be taken into account in order to improve user recommendations. In this regard, more and more methods are emerging that allow recommender systems to adapt to specific contexts [61]. One of these is the Lifetrak system [66], which generates a playlist from the user's music library based on factors related to the user's physical context such as the location, time, day, traffic, level of noise, temperature and weather. Song recommendations in specific time contexts is the objective of a time-aware CF proposal, where micro-profiles are created by using contextual pre-filtering [67]. In this work, different time splits are tested to obtain micro-profiles that best represent the behavior of users during each period. A dataset from Last.fm was used to validate the proposed method. In [68], emotional states are detected by using information inferred from the context of a user at a given time. The contextual recommendation of songs is based on those emotional states. A mood based recommendation of music for videos is proposed in [69]. Both videos and music are projected onto a latent emotional space and a latent factor model is used to find the relevance between videos and music.

The time factor is also the target of some context-aware music recommender systems. In [70], music preferences are dynamically adapted to the contextual changes associated with the users. A sequential pattern algorithm is used to capture contextual states. In [71], changes in user preferences are modeled using hidden Markov models. The sequence of items obtained from the user is used to obtain the model, whose hidden variables represent the context of the user. The context prediction is taken to generate the recommendations. With this proposal, which is validated with data from music playlists, it is possible to improve the accuracy and diversity of the results provided using other methods. Wang et al. [72] proposed a sequence-based method for context-aware music recommendations that combines the general and contextual preferences obtained from users' previous listening sessions.

Deep learning algorithms have also recently been used in context-aware recommender systems. In [73], a convolutional neural network (CNN) is applied in a recommendation approach that considers the time context. The contextual information is part of the input to the network as well as the ratings and the user and item features. The method is validated against other basic time-aware methods with the MovieLens-1m dataset. Only two time contexts are considered in the experiments: weekend and weekday.

Despite the large amount of work published on context-aware systems, there are few studies that have examined time as a contextual variable. In addition, and as far as we know, there are no other studies in which the temporal context and the evolution of preferences are addressed in combination.

## 3. Modeling the Time Effect on Music Recommendations

### 3.1. Problem Statement

Although the effects of time on recommender systems have been the focus of many studies, those addressing particular issues related to the music domain are very few. Most of these particularities are derived from the way users consume musical items, which is different from other products, as well as from the listening habits of users regarding the time context. In this work, both aspects are addressed by means of a recommendation approach incorporating time factors, which is specifically designed for the music area when explicit ratings are unavailable.

In this context, analyzing the evolution of user preferences requires new approaches, since methods defined for other domains are not applicable for music recommendation.

Let us consider a set of users $U$ and a set of songs $S$ where $u_i \epsilon U, i = 1, \ldots, n$ and $s_j \epsilon S, j = 1, \ldots, m$ represent a user and a song, respectively. When explicit ratings are available, each user $u_i$ has a list of ratings that he/she has given to a subset of songs $S_{u_i}$, where $S_{u_i} \subseteq S$. The ratings are stored in a $m \times n$ matrix called the rating matrix, where each element $r_{ij}$ is the rating that a user $u_i$ gives to a song $s_j$.

In the context of this work, explicit ratings are not available, so the methods described in Section 2.2.1 cannot be applied. As seen in that section, both recency-based models and models using time as an independent variable require the use of timestamped ratings. For certain types of items, those typically consumed only once by the user (for instance, a book), these ratings can be obtained from the time spent by the user examining an item or from the item's purchase record. This case corresponds to the one-class CF scenario where $r_{ij}$ can only take the binary values corresponding to the positive and unknown user–item feedback, respectively. However, for music items, which are consumed many times by the user (a song is usually listened to many times by a user), it is very difficult to derive timestamped ratings in an implicit way. Using the number of plays is a common way to obtain this information, which requires a specific time frame that is wide enough to get reliable ratings. Thus, ratings are not associated with a specific time point, but rather with a longer period of time. This situation does not allow, however, the previously described approaches, such as time-SVD++ or decay functions, to be applied. In order to overcome these problems, we propose that the evolution of user preferences be modeled in the same process employed to compute implicit ratings, which are inferred from the plays of songs by users. For this purpose, the concepts of decay play and implicit decay ratings are introduced. The definition of these concepts is given below.

**Definition 1.** *(decay play). Each user $u_i \epsilon U$ has played a subset of songs where each song $s_j \epsilon S$ is played at a known given time $t$. Each play $play_{i,j,t}$ is assigned a weight, called decay play ($dPlay$), which meets the following condition: $dPlay(play_{i,j,t}) < dPlay(play_{i,j,t'})$, when $t' > t$. This property allow us to establish a sequence of decay plays per user and song that is defined in the form: sequence $\{dPlay(u_i, s_j, p)\} \mid p = 1, \ldots, n_p(u_i, s_j)$ and $dPlay(u_i, s_j, p) < dPlay(u_i, s_j, p + 1)$, where $n_p(u_i, s_j)$ is the number of times that the user $u_i$ plays the song $s_j$.*

**Definition 2.** *(implicit decay rating). A rating inferred from the frequency of decay plays ($dPlayF$). This frequency is computed for each pair $(u_i, s_j)$:*

$$dPlayF(u_i, s_j) = \frac{\sum_{p=1}^{n_p(u_i, s_j)} dPlay(u_i, s_j, p)}{\sum_j \sum_{p=1}^{n_p(u_i, s_j)} dPlay(u_i, s_j, p)} \tag{11}$$

This information is represented in the decay play matrix $\mathbf{D} := d_{i,j}$ where $\mathbf{D} \epsilon M_{n \times m}(\mathbb{N})$ and $d_{i,j} := dPlayF(u_i, s_j)$.

The songs played by each user $u_i$ are ordered by their $dPlayF$ values in a sequence defined as sequence $\{dPlayF_{k'}(u_i)\} \mid dPlayF_{k'}(u_i) > dPlayF_{k'+1}(u_i)$, where $dPlayF_{k'}(u_i)$ denotes the decay frequency of a song $s_j$ with rank $k'$. $k' = 1$ is assigned to the song having the highest frequency. Then, the decay rating $r_{i,j}$ for a song with rank k is computed as a linear function of the frequency percentile. These two definitions form the basis of the method proposed in this paper for dealing with the evolution of user preferences when no explicit ratings are available. Managing time context also presents some drawbacks in the music field, especially when the computation of implicit ratings is necessary. One of the main problems of context-aware methods is the sparsity, which is caused by the addition of contextual dimensions. Even in context-free scenarios, there is a certain degree of sparsity because many elements $r_{ij}$ of the rating matrix are null. This is caused by the fact that the number of songs rated by the user is much lower than the number of songs in the system.

When contextual pre-filtering is applied, the time dimension is restricted to a range of values $S_t$ and only the ratings associated to that specific time period are considered to make recommendations for that temporal context. This accentuates the sparsity problem. For instance, to recommend a song to be listened to in the morning, only the ratings corresponding to songs played in the morning can be used if contextual pre-filtering is used. The problem becomes worse when, besides the contextual periods, we have to establish additional periods to calculate the ratings from the number of plays. Moreover, user listening habits regarding the time context are not clearly defined and can vary considerably from one user to another. Therefore, user time-dependent behaviors would be managed in a different way to improve recommendations. In our work, the similarity between users is not only obtained from the ratings they give to songs, but also from the time the songs are played.

*3.2. Time-Aware Music Recommendation Method*

This section is dedicated to presenting the approach proposed to address both the evolution of user preferences and time context in the music recommendation area.

3.2.1. Modeling the Evolution of Preferences

The recency models described in Section 2.2 assume that user preferences change over time in such a way that the more recent the rating is, the more important it is. Usually, an exponential function of time is used to reduce monotonically ratings' importance, but, as previously stated, it requires ratings to be stamped with a date. In the context of this work, explicit ratings are not available; therefore, implicit ratings must be computed from the frequency of plays. Given that the information about the date and time of the plays is available, it can be used to calculate the degree of importance of the implicit ratings obtained from them. Therefore, our proposal can be considered a recency approach, as it involves changes in preferences over time. However, it differs from the models described in Section 2.2 in the fact that it does not require timestamped explicit ratings since the evolution of preferences is encompassed in the process for inducing implicit ratings. In this process, unlike other recency methods, the time function is applied to plays instead of ratings.

The first step of the proposal is the calculation of the decay plays defined in the previous section from the timestamped plays. In the same way that the weight of ratings decreases as their age increases, the importance of plays will also decrease. Thus, ratings can be calculated from the frequencies of weighted plays. We propose the use of an exponential function to weight the plays according to the moment in which they occur. As stated before, a decay play (*dPlay*) is defined as the weight that a play of a song $s_j$ by a user $u_i$ has according to its timestamp $t_{i,j,p}$, as seen in Equation (12), with 1 being the maximum weight, which is assigned to the most recent play.

$$dPlay(u_i, s_j, p) = e^{-\lambda(t_{max} - t_{i,j,p})/T_u} \qquad (12)$$

The most recent time is designed as *t*<sub>max</sub>. The parameter $\lambda$ is used to adjust the decay rate. $T_u$ is used to convert the units in which $t_{i,j,p}$ and *t*<sub>max</sub> are measured to the target units, which depend on the application domain. We have used years as the target unit since changes in musical trends are not usually very drastic.

Other functions can be used; however, several studies in which the evolution of ratings has been analyzed, have proved that exponential decay is more suitable. This occurs because the gradient of the curve reflects the change of their importance over time in a more realistic way.

Although this model is indicated for users who have been registered in the system for some time, it can also be applied in the cold-start scenario referring to new users who only have recent plays. In these cases, the value of the decay plays will be 1 or close to 1, so they will receive recommendations very similar to those received without considering the evolution of preferences.

The frequency of weighted plays will be used to obtain implicit ratings that incorporate time effects. The frequency of decay plays for a user $u_i$ and a song $s_j$ within a given time window is computed as indicated in Equation (11).

**Algorithm 1** Computation of decay plays, implicit ratings and the average time of day attribute.

**Input**: $U = \{u_1, u_2, \ldots, u_n\}$
$S = \{s_1, s_2, \ldots, s_m\}$
$t_{i,j,p} \ \forall \ i,j,p \mid i = 1, \ldots, n \ \land \ j = 1, \ldots, m \ \land \ p = 1, \ldots, n_p(u_i, s_j)$
$\lambda, T_u$
**Output**: $R, DT$ // $R$: Rating matrix, $DT$: Vector of average time of day for every user $u_i$

1:  $R := r_{i,j}$ , $R \in M_{n \times m}(\mathbb{N})$
2:  $DT := dTavg_i$ , $DT \in M_{n \times 1}(\mathbb{N})$
3:  $t_{max} := \max(t_{i,j,p})$
4:  **for** $i = 1$ to $n$ **do**
5:     $dPlayFDem(u_i) \leftarrow 0$
6:     **for** $j = 1$ to $m$ **do**
7:        $dPlayFNum(u_i, s_j) \leftarrow 0$
8:        $NP(u_i) = 0$
9:        **for** $p = 1$ to $n_p(u_i, s_j)$ **do**
10:          $dPlay(i,j,p) := e^{-\lambda(t_{max} - t_{i,j,p})/T_u}$
11:          $dPlayFNum(u_i, s_j) \leftarrow dPlayFNum(u_i, s_j) + dPlay(i,j,p)$
12:       **end for**
13:       $dPlayFDem(u_i) \leftarrow dPlayFDem(u_i) + dPlayFNum(u_i, s_j)$
14:       $NP(u_i) \leftarrow NP(u_i) + n_p(u_i, s_j)$
15:    **end for**
16: **end for**
17: **for** $i = 1$ to $n$ **do**
18:    **for** $j = 1$ to $m$ **do**
19:       $dPlayF(u_i, s_j) \leftarrow dPlayFNum(u_i, s_j) / dPlayFDem(u_i)$
20:    **end for**
21: **end for**
22: **for** $i = 1$ to $n$ **do**
23:    $sequence \ \{Freq_{k'}(i)\} := sequence \ \left\{dPlayF(u_i, s_j)_{k'}\right\} \ \forall \ j \mid dPlayF(u_i, s_j) > 0 \ \land \ dPlayF(u_i, s_j)_{k'} > dPlayF(u_i, s_j)_{k'+1}$
24:    **for** $j = 1$ to $m$ **do**
25:       $F(u_i, s_j) \leftarrow 0$
26:       Set $k$ value $\mid Freq_k(i) = dPlayF(u_i, s_j)$
27:       **for** $k' = 1$ to $k - 1$ **do**
28:          $F(u_i, s_j) \leftarrow F(u_i, s_j) + Freq_{k'}(i)$
29:       **end for**
30:       $r_{i,j} \leftarrow 4 \ (1 - F(u_i, s_j))$
31:    **end for**
32: **end for**
33: **for** $i = 1$ to $n$ **do**
34:    $T(u_i) \leftarrow 0$
35:    **for** $j = 1$ to $m$ **do**
36:       **for** $p = 1$ to $n_p(u_i, s_j)$ **do**
37:          $T(u_i) \leftarrow T(u_i) + t_{i,j,p}$
38:       **end for**
39:    **end for**
40:    $dTavg_i \leftarrow T(u_i) / NP(u_i)$
41: **end for**
42: **return** $R, DT$

Once the decay play frequencies are computed, the next step is to obtain the implicit ratings from them. To do this, we recommend following the procedure by Pacula [74], since this method is appropriate for power-low distributions, like in this work, where a higher number of plays is concentrated on a few songs. A rating $r_{i,j}$ for the user $u_i$ and the song $s_j$ is obtained as a linear function of the frequency percentile by using Equation (13), where $dPlayF_{k'}(u_i)$ denotes the decay play frequency of a song with the rank $k'$ in the ranked list of the user $u_i$, as introduced in Definition

2. Simple frequency functions [75] can also be used to transform plays into ratings but are less suitable for these types of situations.

$$r_{i,j} = 4\,(1 - \sum_{k'=1}^{k-1} dPlayF_{k'}(u_i))\circledast \tag{13}$$

The ratings calculated from the frequency of decay plays reflect the evolution over time of user preferences. For users who suffer from the cold-start problem because they have recently joined the system, this evolution will be reflected to a lesser extent and the decay ratings calculated for them will be similar to the ratings provided.

3.2.2. Modeling Time Context

User time-dependent behavior in the music domain is another aspect that can influence recommendations. Users have different habits depending on the time of day when they are listening to music, which can be taken into account to improve recommender models. Context-aware recommendation methods are based on these general patterns of user behavior. They are used to manage situations in which time is a contextual variable, since user preferences may depend on this. Subsequently, the day is divided into time intervals and different recommendations are generated for each of these time windows. As previously mentioned, this can cause the problem of sparsity, which is even worse in the music domain, where implicit ratings are computed from the number of plays due to the need for additional intervals. In the experimental study carried out in this work, in which the previously described method is tested, only two contextual time intervals, morning and evening, are considered as a way to minimize this problem. Additionally, the pre-filtering approach has been applied to generate recommendations for every period.

Considering a tridimensional data space $D$ where the ratings in set $D$ depend on users, items and time context, the contextual pre-filtering reduces the space from three to two dimensions by setting the time to a value $t$, as seen in Equation (14), or a range of values $S_t$, as seen in Equation (15).

$$R_{U \times I \times T}^{D} \rightarrow R_{U \times I}^{D\,[T=t]} \tag{14}$$

$$R_{U \times I \times T}^{D} \rightarrow R_{U \times I}^{D\,[T \in S_t]} \tag{15}$$

where $U$ is the set of users, $I$ is the set of items and $T$ is the set of time values when ratings are given to the items. When the context of time is introduced, $T$ is usually restricted to a specific interval $S_t$, called the contextual segment. After reducing the space to two dimensions, traditional CF methods can be applied to every contextual segment. For instance, the recommendation of songs to be played in the morning only use data belonging to the morning contextual segment.

3.2.3. Modeling User Listening Behavior

In addition to establishing contextual time intervals, we also analyze the listening habits of users throughout the day. In this case, we focus on the individual behavior patterns of each user to achieve a greater customization of the recommendations. This data analysis reveals very different behaviors among users. Nevertheless, certain likenesses among users who play songs at similar times of day can be found. Moreover, in order to introduce listening habits in the recommender models, it is necessary to define a variable that represents this behavior. The average time of day for every user $u_i$ can be a suitable factor, which is defined below.

$$dTAvg_i = \frac{\sum_j \sum_p t_{i,j,p}}{NP_i} \tag{16}$$

where $t_{i,j,p}$ is the time of day for the play $p$ of the song $s_j$ by the user $u_i$ and $NP_i$ is the total number of plays of the user $u_i$.

This allows us to use the contextual variable of the time of day as an additional attribute for computing the similarity between users, which may ultimately improve the recommendation reliability. Once this new similarity $sim(u_a, u_i)$ between the active user $u_a$ and the other users $u_i$ is computed, Equation (3) can be used to predict the rating that the active user would give to a song $s_j$.

We use the cosine function to compute the user similarity by taking ratings and $dTAvg$ as attributes. In general, for any number of attributes, if $d$ is the number of dimensions corresponding to the considered attributes, $X_a = (x_{a,1}, x_{a,2} \ldots x_{a,d})$ a vector containing the values of the attributes for the active user $u_a$, and $X_i = (x_{i,1}, x_{i,2} \ldots x_{i,d})$ a vector containing the values of the attributes for the user $u_i$, then the cosine similarity between $u_a$ and $u_i$ is defined as follows:

$$cosSim(u_a, u_i) = \frac{\sum_{z=1}^{d} x_{a,z} \, x_{i,z}}{\sqrt{\sum_{z=1}^{d} x_{aj}^2} \sqrt{\sum_{z=1}^{d} x_{ij}^2}} \tag{17}$$

The procedure for computing the implicit ratings from decay play frequencies, as well as the time of day variable, is described by Algorithm 1. This algorithm provides all the variables necessary as an output to apply collaborative filtering methods according to our proposal. In the experimental study detailed in the following section, the procedure is applied to the context-free dataset, as well as to the subsets corresponding to different contextual intervals regarding the time of the day.

## 4. Experimental Study and Validation

An experimental study using data collected by Oscar Celma (https://www.upf.edu/web/mtg/lastfm360k) from Last.fm was conducted to validate the proposed method. The dataset used in this study consists of more than 80,000 songs listened to by 50 users over a 2-year period, creating a collection of more than 420,000 timestamped plays. The validation metrics used in the study are described in Section 4.1 and the baseline methods are described in Section 4.2.

The first part of the study, presented in Section 4.3, is dedicated to validating the method for obtaining the implicit ratings considering the evolution of user preferences. In this part, two types of experiments are reported. First, the proposed method was applied without taking into account any contextual information. Then subsequently, the context of the time of day was considered and the results were compared.

The second part, included in Section 4.4, is dedicated to introducing a time variable related to the listening habits of users, in an attempt to characterize their behavior. This variable was incorporated into collaborative filtering as an additional attribute for improving the recommendations. The work in this part is carried out in the same manner as mentioned above, and the results obtained are also compared in this section.

Given that it is assumed that user preferences change over time, the validation was performed by splitting the entire dataset into a training set containing the data corresponding to the first 15 months and a test set including data from the last 2 months. The implicit ratings were computed. The data were divided before carrying out the rating computation, so that the ratings and the decay ratings could be computed, respectively, from the number of plays and decay plays in each subset separately by means of Equation (13). Their values were in the range (0, 4). The same procedure was used when contextual periods were considered, computing both types of ratings separately for each training and test set for each contextual period. This is the most suitable way of validating recency-based models [33]. Each record in the datasets consists of the user ID, song ID and either the calculated rating or decay rating for the pair user-song, depending on the approach being evaluated. Therefore, the ground truth used to calculate the error metrics for ratings and decay ratings only varies in these implicit ratings, while the user-song pairs were the same. In addition, the test set records containing user-song pairs, which were also contained in the training set records, were removed. This is done because the recommendation list cannot contain songs that have been previously played by the user, as stated in CF. We used the same datasets for both the rating prediction and top-N recommendations.

The validation strategy followed for top-N recommendation is AllItems [76], in which the entire set of items except those rated by the user in the training set, is used to make the recommendation list. This is considered a suitable strategy to compare the performance of the algorithms. TestRatings is an alternative method that only considers the items in the test set that have ratings of the target user to generate the top-N list for that user. Therefore, the evaluation is only performed over a reduced number of items of known relevance, which in practice is not a realistic scenario. AllItems adds some items to the list which were non-relevant since they have no rating for a given user in the test set. Thus, it can give lower precision values than TestRatings but it is expected that this fact will similarly affect all recommendation algorithms. Consequently, this procedure is more suitable for comparing algorithms since it reflects a more realistic scenario [76].

*4.1. Validation Metrics*

The proposed method is validated for both rating predictions and top-N recommendations. The metrics used to evaluate the rating prediction reliability were the MAE (mean absolute error), RMSE (root-mean-square error) and NMAE (normalized mean absolute error) [77]. For top-N recommendations, the following ranking specific metrics were used: AUC (area under the receiver operating characteristics (ROC) curve), NDCG (normalized discounted cumulative gain), MAP (mean average precision), and precision for the first 5, 10 and 15 items of the top-N list (prec@5, prec@10 and prec@15). Most of these measures are defined in the information retrieval area [78]. The metrics used in the validation are defined below:

MAE is the average of the absolute error or difference between the actual rating (*r*) and the predicted rating (*pr*).

$$MAE = \frac{1}{n}\sum_{i=1}^{n} |r_i - pr_i| \quad (18)$$

where $n$ is the number of predictions, $r_i$ is the actual rating and $pr_i$ is the predicted rating.

The NMAE is obtained by normalizing the MAE with respect to the range of rating values. Thus, it is independent of the evaluation range.

$$NMAE = \frac{MAE}{r_{max} - r_{min}} \quad (19)$$

The RMSE is a good measure to compare the errors of different predictive models. It represents the sample standard deviation of the differences between the predicted values and observed values.

$$RMSE = \sqrt{\frac{1}{n}\sum_{i=1}^{n}(r_i - pr_i)^2} \quad (20)$$

The metrics used in the validation of the top-N recommendations are described below.

The AUC is a metric obtained from the representation of correct and incorrect prediction rates, in particular, the true positive rate (TPR) and false positive rate (FPR). To compute them, it is necessary to define other variables: true positives (TPs) are positive instances correctly classified; false positives (FPs) are negative instances classified as positive; true negatives (TNs) are negative instances correctly classified; false negatives (FNs) are positive instances classified as negative.

$$TPR = TP/(TP + FN) \quad (21)$$

$$FPR = FP/(TN + FP) \quad (22)$$

The ROC (receiver operating characteristics) curve is the representation of the $TPR$ against the $FPR$ of classifiers induced by the same algorithm and from the same data but with different probability threshold settings. Point (0,0) of the ROC graph corresponds to a classifier that classifies all examples as negatives, and point (1,1) corresponds to a classifier that classifies all examples as positive. The best learning system will be the one that provides a set of classifiers with a greater area under the ROC curve (AUC). The larger the area, the more distant the ROC curve is from the line joining the two points, which represents the random classification.

The NDCG is a measure based on the assumption that the lower the ranked position of a relevant item, the less useful it is for the user. When applying this metric, the gain is accumulated starting at the top of the ranking. The graded relevance value of items in lower positions is reduced in a quantity logarithmically proportional to the position of the result. DCG is the discounted cumulative gain accumulated at a particular rank *k*.

$$DCG_k = \sum_{i=1}^{k} \frac{r_i}{log_2(i+1)} \quad (23)$$

$$IDCG_k = \sum_{i=1}^{|REL|} \frac{r_i}{log_2(i+1)} \quad (24)$$

where |REL| is the number of the best ratings up to position *k*.

$$NDCG_k = \frac{DCG_k}{IDCG_k} \quad (25)$$

The MAP is the average of the precision value obtained from all users and items in the top-N lists, where prec@K is the proportion of recommended items in the top-k set that are relevant. Since the rank-based metrics for validating top-N recommendations are referred to a list with a given size (N), the relevant items are the N items with the highest implicit rating values in the test set. The top-N lists generated from the predicted ratings will contain a proportion of such relevant items and that value constitutes the precision of each list.

*4.2. Baseline Methods*

The baseline methods used in the comparative study belong to the two most used categories in the field of recommender systems: memory-based methods and matrix factorization. The former usually give very good results when there are no sparsity problems while matrix factorization methods, although they are also affected by the dispersion, usually perform better with sparser data. High degrees of dispersion generally occur when users have to explicitly express their preferences, as in such cases the number of ratings for the products to be recommended is usually small. In this work, the proposal to calculate the ratings from the plays of songs allows for a greater number of ratings and minimizes the sparsity problem. In this scenario, it is difficult to predict which approach will provide better results. In any case, our aim is not to compare these methods but to prove that, on the one hand, they perform better with decay ratings than with ratings, and, on the other hand, that our proposal of incorporating user listening behaviors outperforms all the baseline models.

The specific methods tested for rating predictions were the user-based K-nearest neighbor (K-NN), matrix factorization (MF), biased matrix factorization (BMF) [79] and factor wise matrix factorization (FWMF) [80]. K-NN was tested using both the cosine similarity and Pearson correlation coefficient. The implementation of those algorithms provided by the RapidMiner tool (https://rapidminer.com) were used. The number of k neighbors was set to 15, since higher values do not improve the results. Other methods for modeling the evolution of user preferences cannot be applied since they need timestamped ratings, as explained in detail in Section 3.

The methods applied for providing top-N recommendations were the user K-NN, weighted regularized matrix factorization (WRMF) [32,81] and Bayesian personalized ranking matrix factorization (BPRMF) [31].

*4.3. Modeling Temporal Evolution of Preferences*

4.3.1. Rating Prediction

The first experiments were carried out without considering the context of time. Thus, the entire dataset of song listening information taken within a 24-h period was used. We compared the outcome of several recommender methods using implicit ratings, computed following the method proposed by Pacula, in two ways, without considering temporal effects (ratings), and considering those effects by applying our proposed method (decay ratings).

The decay ratings were computed using Equations (11–13). The optimal value of $\lambda$ in Equation (11) was set to 0.4 after some preliminary tests proved that this value yields the lowest error rates.

The first results shown in Table 1 correspond to the rating prediction for the complete day when the time effect was not considered in the generation of implicit ratings (ratings) and when it was taken into account according to our proposal (decay ratings). We can observe that a significant reduction in the rate of errors (columns "improv.") was achieved when using decay ratings against the ratings of all recommendation methods. The MAE was reduced by 8.1–22.5%.

The next step was to validate the method when considering the context of the time of day. To do so, the dataset was divided into time intervals according to the proposal of [61], where time intervals were established for the morning and evening hours. Morning was defined as the hours between 5 am and 6 pm and evening as the remaining hours of the day. Baltrunas and Amatriain [61] study the time partition that provides the lowest error rates using a Last.fm dataset for artist recommendations. However, our aim is not to prove that the context of time can improve recommendations, but to prove that the improvements of our approach (using decay ratings instead of ratings), achieved without considering the time context, are extensible to contextual time intervals.

Baltrunas and Amatriain report an improvement of 2.4% in the MAE associated with recommendations for the morning and evening time frames as opposed to the complete 24-h period. The dataset was extracted from Last.fm and used to recommend artists to users by means of the CF method. A five-fold cross validation was applied in the experiments. In a preliminary study, we applied the same method and the same validation procedure to our dataset, in this case for recommending songs. However, instead of decreasing, the values for the MAE for the morning and evening periods slightly increased, when compared to the complete 24-h period. Following on, we proceeded to check the behavior when using the previously described training and test set partition, in which examples of the test set correspond to the dates subsequent to those of the training set examples.

Table 1 also presents the results obtained in the experiments involving the two time periods (morning and evening). An analysis of the values showed that the most significant result is that the error rates for the morning and evening periods were substantially lower than those of the complete time period (day) when traditional ratings were used. However, this reduction was less obvious when the decay ratings were used. For the ratings, the MAE decreases between 16.3% and 18.8% when comparing the morning period with the complete day, and between 6.5% and 17.1% when comparing the evening with the complete day. For the decay ratings, the reduction in the MAE varied between 0.6% and 14.6% when comparing the morning period with the complete day and between 0.6% and 6.7% when comparing the evening with the complete day. In the latter case, an increment of 2.1% was produced for K-NN with the cosine similarity, which was the method that yielded the worst results in all the experiments.

Another conclusion that can be drawn from this analysis is that the decay ratings provided the best results for all methods tested in this study. Although, for the morning and evening periods, the improvement was not as great as the one achieved for the complete day, where the reduction in the MAE varied between 1.1% and 11.3%.

Figure 1 shows a comparison of the results obtained in these experiments, where it can be seen that the error rates (MAE and RMSE) are lower for the decay ratings as compared to the ratings, obtaining the longest distance for the complete day. The best values for each method and the percentages of improvement are highlighted in bold.

**Table 1.** The root-mean-square error (RMSE), mean absolute error (MAE) and normalized mean absolute error (NMAE) values obtained for the rating prediction when contextual information is not considered (day) and when contextual pre-filtering is carried out (morning and evening).

|  |  | MAE | | | RMSE | | | NMAE | |
|---|---|---|---|---|---|---|---|---|---|
|  |  | Ratings | Decay Ratings | Improv. | Ratings | Decay Ratings | Improv. | Ratings | Decay Ratings |
| Day (24 h.) | K-NN Pearson | 1.265 | **1.037** | **18.02%** | 1.570 | **1.299** | **17.26%** | 0.316 | **0.259** |
|  | K-NN Cosine | 1.303 | **1.066** | **18.19%** | 1.601 | **1.340** | **16.30%** | 0.326 | **0.267** |
|  | MF | 1.263 | **1.028** | **18.61%** | 1.569 | **1.291** | **17.72%** | 0.316 | **0.257** |
|  | BMF | 1.229 | **0.952** | **22.54%** | 1.529 | **1.179** | **22.89%** | 0.307 | **0.238** |
|  | FWMF | 1.279 | **1.176** | **8.05%** | 1.582 | **1.499** | **5.25%** | 0.320 | **0.294** |
| Morning | K-NN Pearson | 1.048 | **1.006** | **4.01%** | 1.310 | **1.273** | **2.82%** | 0.262 | **0.251** |
|  | K-NN Cosine | 1.090 | **1.060** | **2.75%** | 1.353 | **1.345** | **0.59%** | 0.273 | **0.265** |
|  | MF | 1.025 | **0.948** | **7.51%** | 1.260 | **1.186** | **5.87%** | 0.256 | **0.237** |
|  | BMF | 1.008 | **0.919** | **8.83%** | 1.240 | **1.147** | **7.50%** | 0.252 | **0.230** |
|  | FWMF | 1.055 | **1.004** | **4.83%** | 1.294 | **1.271** | **1.78%** | 0.264 | **0.251** |
| Evening | K-NN Pearson | 1.094 | **1.023** | **6.49%** | 1.370 | **1.293** | **5.62%** | 0.274 | **0.256** |
|  | K-NN Cosine | 1.100 | **1.088** | **1.09%** | 1.389 | **1.371** | **1.30%** | 0.275 | **0.272** |
|  | MF | 1.047 | **0.978** | **6.59%** | 1.317 | **1.232** | **6.45%** | 0.262 | **0.245** |
|  | BMF | 1.034 | **0.946** | **8.51%** | 1.294 | **1.183** | **8.58%** | 0.258 | **0.237** |
|  | FWMF | 1.196 | **1.097** | **8.28%** | 1.527 | **1.399** | **8.38%** | 0.256 | **0.227** |

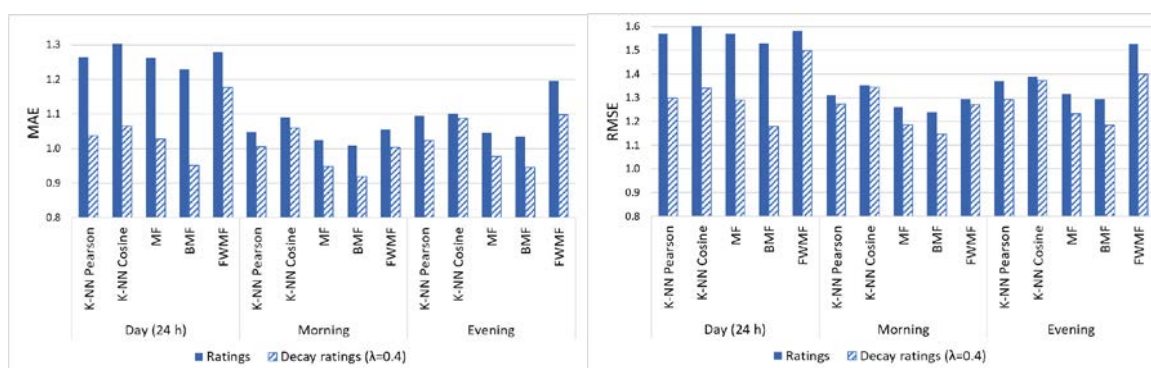

**Figure 1.** The MAE and RMSE values given by some recommendation methods using ratings and decay ratings for different time periods.

Finally, to confirm that the method can be applied in the col-start scenario, as stated in the description of the proposal, we have conducted some additional experiments. From the entire dataset, the five most recent users have been selected since they have a much shorter play history than the oldest users. To perform this validation, records for all except the five selected users were removed from the test set. The objective was to prove that the use of the decay ratings with the recommendation methods is better or at least equal to the use of the ratings for new users. The study has been carried out only for a 24-h period because when doing the contextual pre-filtering for short periods with only five users, the test set would have been a very small size. The results are shown in Table 2. The best values for each method in the comparison of ratings and decay ratings and the percentages of improvement are highlighted in bold.

We can observe that the results obtained with decay ratings are better than those obtained with ratings for all recommendation methods, although the percentage of improvement was lower than when using the complete data set. These results were predictable since the evolution of preferences of users who have been registered in the system for a shorter period of time, is reflected to a lesser extent. Therefore, decay ratings and ratings have similar values for these users.

**Table 2.** The RMSE, MAE and NMAE values obtained for the rating prediction in the cold-start scenario.

|  | **MAE** | | | **RMSE** | | | **NMAE** | |
|---|---|---|---|---|---|---|---|---|
|  | Ratings | Decay Ratings | Improv. | Ratings | Decay Ratings | Improv. | Ratings | Decay Ratings |
| K-NN Pearson | 0.898 | **0.890** | 0.89% | 1.162 | **1.153** | 0.77% | 0.225 | **0.222** |
| K-NN Cosine | 0.985 | **0.948** | 3.76% | 1.228 | **1.212** | 1.30% | 0.246 | **0.237** |
| MF | 0.788 | **0.777** | 1.40% | 0.999 | **0.998** | 0.10% | 0.197 | **0.194** |
| BMF | 0.785 | **0.775** | 1.27% | 0.977 | **0.982** | 0.51% | 0.196 | **0.194** |
| FWMF | 0.803 | **0.792** | 1.37% | 0.995 | **0.993** | 0.20% | 0.201 | **0.198** |

4.3.2. Top-N Recommendation

After analyzing the rating prediction results, the reliability of the top-N recommendations was studied. This kind of recommendation consists of a ranked list containing the N songs with the highest values of predicted ratings. The evaluation of these lists was usually performed by means of rank-based metrics. We used N=100 for obtaining the results.

The results of the tests performed using ratings and decay ratings are presented in Table 3 and Figure 2. The table shows the values of the AUC, NDCG and MAP for the experiments corresponding to the complete day, morning and evening (the best results are shown in bold). The figure illustrates the comparison of the two types of evaluation (ratings and decay ratings) according to these metrics and others such as prec@5, prec@10, prec@15.

**Table 3.** The area under the receiver operating characteristics (ROC) curve (AUC), normalized discounted cumulative gain (NDCG), and mean average precision (MAP) values obtained for the rating prediction when contextual information is not considered (day) and when contextual pre-filtering is performed (morning and evening).

|  |  | **AUC** | | **NDCG** | | **MAP** | |
|---|---|---|---|---|---|---|---|
|  |  | Ratings | Decay Ratings | Ratings | Decay Ratings | Ratings | Decay Ratings |
| Day (24 h.) | K-NN | 0.422 | 0.422 | 0.327 | 0.327 | 0.006 | 0.006 |
|  | WRMF | 0.500 | **0.525** | 0.331 | **0.336** | 0.007 | **0.009** |
|  | BPRMF | 0.528 | **0.535** | 0.330 | **0.332** | 0.005 | **0.006** |
| Morning | K-NN | 0.413 | 0.413 | 0.315 | 0.315 | 0.005 | 0.005 |
|  | WRMF | 0.516 | **0.519** | 0.316 | **0.321** | 0.006 | **0.008** |
|  | BPRMF | 0.473 | **0.491** | 0.312 | **0.313** | 0.004 | 0.004 |
| Evening | K-NN | 0.378 | 0.378 | 0.297 | 0.297 | 0.004 | 0.004 |
|  | WRMF | 0.590 | **0.597** | 0.316 | **0.322** | 0.009 | **0.011** |
|  | BPRMF | 0.562 | **0.539** | 0.306 | **0.311** | 0.004 | **0.006** |

When the K-NN was applied, the values of the AUC, NDCG and MAP were the same for the decay ratings and ratings in spite of the prediction that the preferences would be better when using decay ratings against ratings, as shown in Table 1. After a detailed examination of the predictions and the top-N lists generated for some users, we verified that the predictions of decay ratings were closer to the actual values than the prediction of ratings, but the ordered list of songs provided by the algorithm is very similar. This indicates that, despite obtaining better results in the prediction when using decay ratings, these do not translate into a great change in the order of the songs in terms of their relevance. Thus, all rank-based metrics (those used for top-N recommendations) give the same results for K-NN. This behavior may be produced by the characteristics of the cosine similarity metric involved in this method. Since the similarity is based on the angle between two vectors, this measure is more sensitive to the fact that two users have ratings on the same items than to the magnitude of those ratings. This is especially important in problems such as the one addressed in this paper, where there are no negative ratings, since it is assumed that users play the songs they like.

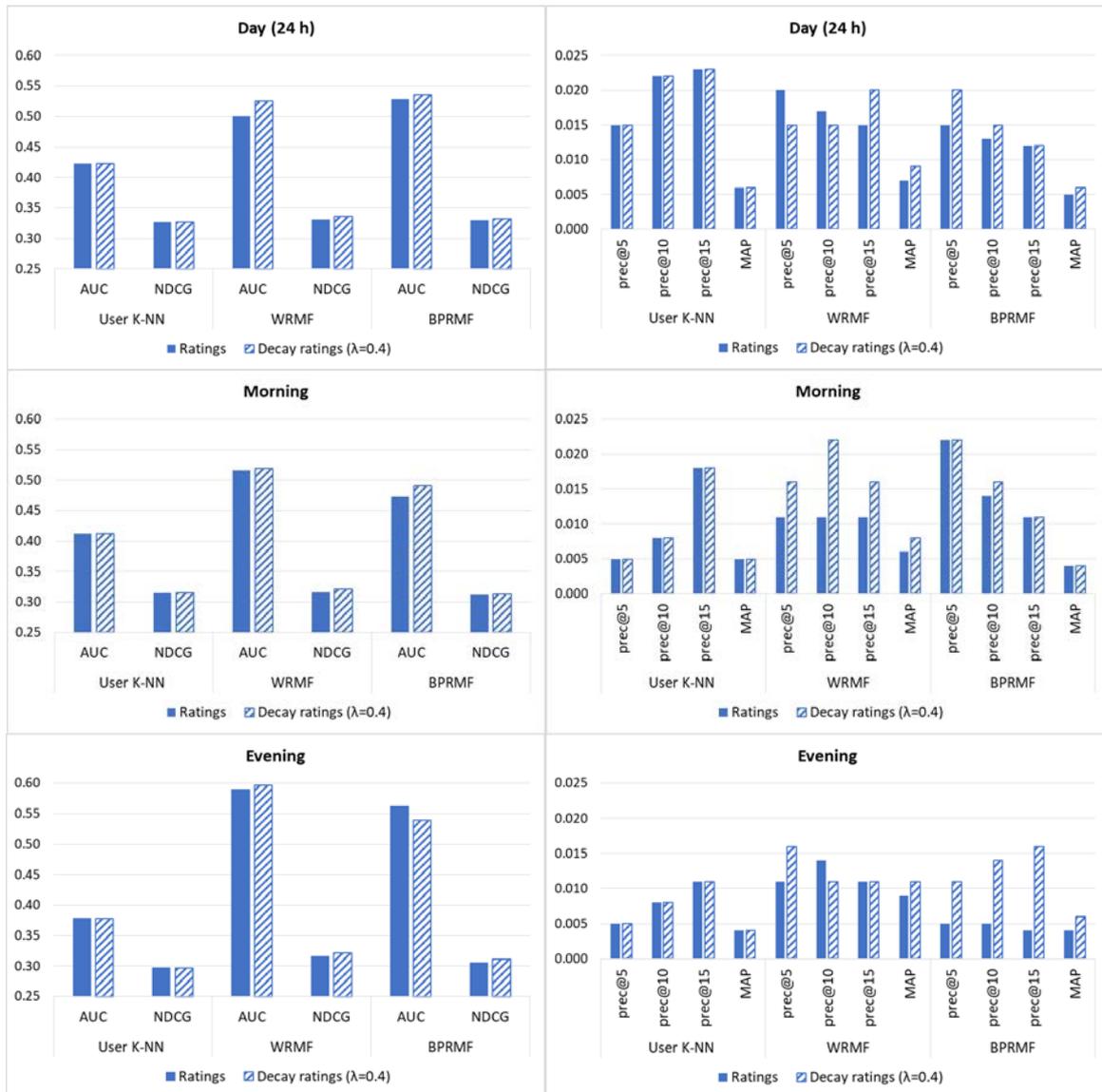

**Figure 2.** Values of the top-N metrics given by user K-nearest neighbor (K-NN), weighted matrix factorization (WRMF) and Bayesian personalized ranking matrix factorization (BPRMF) using rating and decay ratings.

When WRMF and BPRMF were used, the values of the AUC and NDCG were higher for the decay ratings than for the ratings (except the AUC in the evening periods with BPRMF), although the improvements were not as good as those of preference predictions reported in Table 1. The MAP for the decay ratings was higher or equal than for the ratings. The behavior for prec@5, prec@10 and prec@15 was similar for both kinds of evaluations when using the K-NN. However, for WRMF and BPRMF, the decay ratings performed better than the ratings, except for prec@5 and prec@10 for the complete day and prec@5 for the evening period. We can also find an explanation for this performance of the rank-based metrics through the detailed analysis of the individual results. We have verified that some of the most relevant songs for users have very similar ratings and/or decay ratings. Hence, small fluctuations in the predicted values can cause a change in the order of the songs in the top-N list and make an item go out or into the top-5, top-10 or top-15 lists from which these metrics are computed. In some cases, this fact leads to some of the rank-based metrics being worse for the decay ratings than for the ratings, even though the predictions of the decay ratings are better than the predictions of the ratings. Additionally, the greatest differences between the lists produced with the ratings and decay ratings occur for ranks higher than 15, so these changes are not reflected in these metrics. The behaviors described above may be aggravated by the high number of songs available for recommendation, which makes it difficult to obtain a rank close to the actual rank, since little changes in the values of ratings may change the position of the songs in the list.

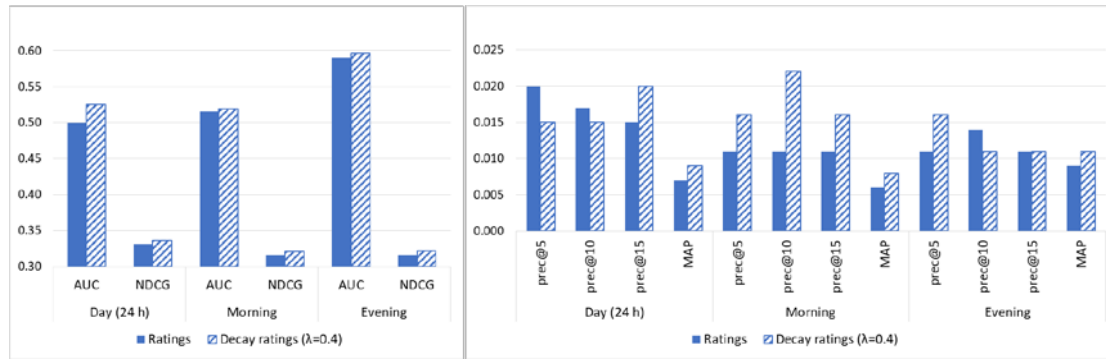

**Figure 3.** Values of the top-N metrics given by WRMF using the ratings and decay ratings for different day time periods.

In Figure 3, the results obtained for the complete day and the morning and evening periods are compared. Only the outcome of the WRMF method with ratings and decay ratings is presented, since the user K-NN produced worse results and did not give appreciable differences between the ratings and decay ratings. The results of the metrics the AUC, NDCG and MAP do not reveal a clear improved performance when the morning and evening time periods are considered, compared to the complete day. The NDCG value is always lower for the morning and evening time periods, while the AUC and MAP values are higher for the evening period when using both the ratings and decay ratings. However, for the morning period, the AUC was higher only when using ratings, and the MAP was lower for ratings and decay ratings when compared to the complete day. Regarding prec@5, prec@10, prec@15, the behavior is irregular, similar to what is observed for the rating prediction.

In general, it can be seen that all the precision values are low, regardless of the time period, the type of ratings and the recommendation method. This fact is compatible with the reasoning given at the beginning of Section 4 regarding the method used to generate the top-N lists, which is suitable for comparison between methods but not for evaluating the performance in absolute terms. The high number of songs in the dataset aggravates this problem.

### 4.4. Modeling Listening Habits of Users

In the second part of the work, we studied the frequency of plays throughout the day to detect the listening behavior of users in relation to time. We analyzed the play frequencies for all users as a complete group (Figure 4), as well as for each individual user (Figures 5 and 6).

Figure 4 shows the distribution of the frequency values during the morning intervals and the distribution of values during the evening intervals, according to the partition mentioned in the previous section. Therefore, this figure shows the listening pattern of all users in the dataset over the day, from 0 to 24 h. We can observe that the maximum frequency was reached around 6 p.m. The figure also shows that morning frequencies present some fluctuations, while the evening frequencies gradually decay during 6 pm to 5 am. When time intervals are established to make context-aware recommendations, this overall pattern is exploited. However, in order to achieve a greater degree of personalization, it is necessary to take into account the individual patterns of each user.

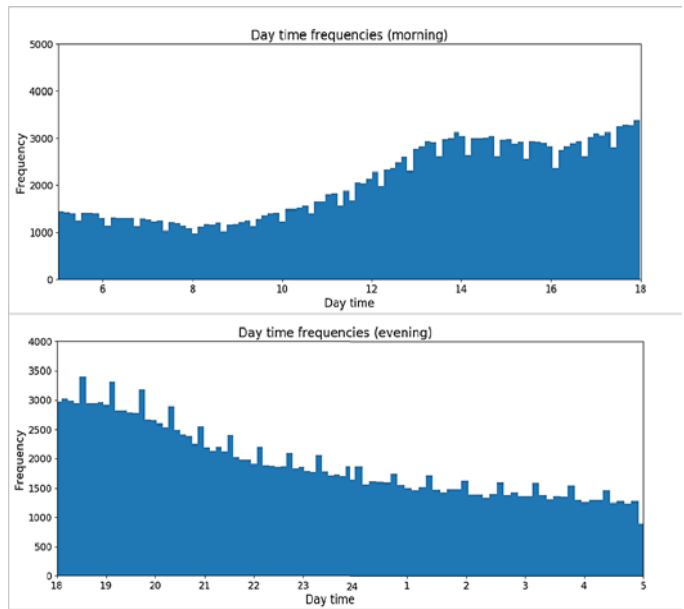

**Figure 4.** Day time frequencies of plays during the morning intervals and the evening intervals.

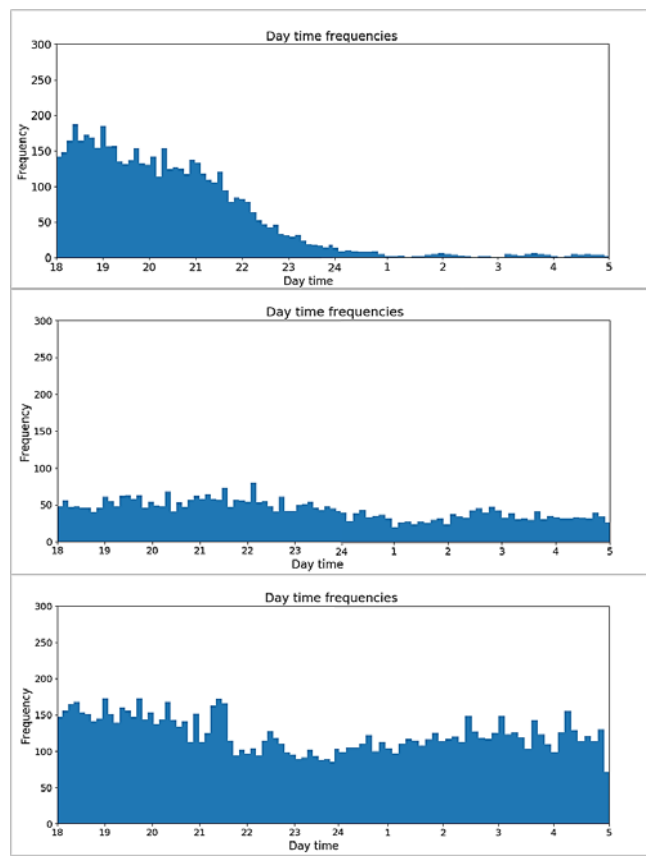

**Figure 5.** Individual day time frequencies of plays for different users during the evening intervals.

When the play frequencies were analyzed individually, different distributions could be found for the various users, since their listening habits were not the same. Figures 5 and 6 present a sample of users exhibiting different behaviors for both the evening and morning intervals. We can see there was greater variability during the morning period than in the evening, at which time the distributions were more similar.

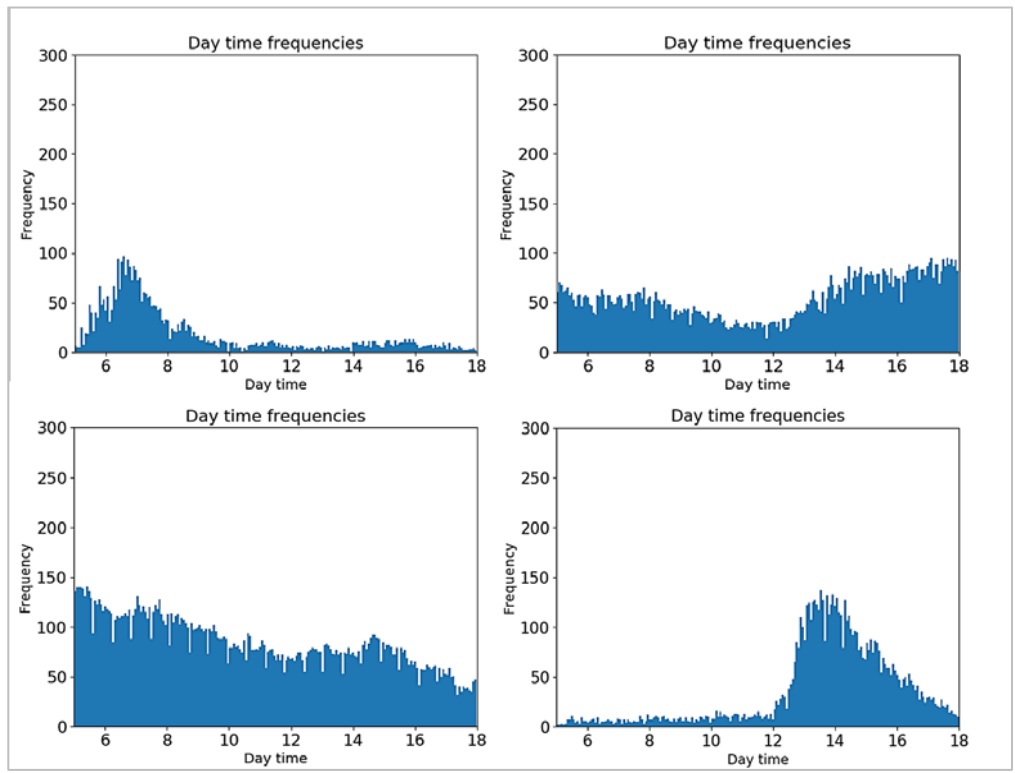

**Figure 6.** Individual day time frequencies of plays for different users during the morning intervals.

Based on this analysis, we propose to use the time listening behavior to characterize the users and to incorporate this behavior in the recommender models as a means to improve the reliability of the recommendations.

The average time of day for each user $u_i$ was used to introduce the listening habits in the recommendation model, which was managed as an additional attribute that allows the similarity to be detected between users in collaborative filtering. The incorporation of this variable into collaborative filtering was tested for both the rating prediction and top-N recommendations. The method used was the K-nearest neighbor (K-NN) user-based collaborative filtering, which is called "user attribute K-NN" and is when other attributes besides the ratings are used to compute the similarity between users. In this case, the metric used is the cosine-similarity. Since the additional attribute used in the K-NN captures the time-dependent user listening behavior, the proposed method was named "user-time K-NN".

4.4.1. Rating Prediction

The user-time K-NN method, which incorporates the time variable, was tested for the complete day and the morning and evening intervals using classical ratings, as well as the decay ratings defined in this work, where $\lambda$ was set to 0.4. Table 4 shows the RMSE, MAE and NMAE values obtained in the tests. As in the tables above, the values in bold represent the best results. Figure 7 presents a comparison of the error rates, the MAE and RMSE, of all methods tested in this study applied to the complete day dataset, as well as to the morning and evening intervals. The last bars of each group represent the results obtained using the user-time K-NN proposal. In all the time periods studied, the introduction of the new attribute, which captures the users' listening habits throughout the day, lead to lower error rates, as can be seen in the graphs. By comparing the results of the user-time K-NN with BMF, the second-best method, the reduction in the MAE is 2.2%, 1.8% and 0.3% for the complete day, morning and evening, respectively, when using ratings. In the case of the decay ratings, the reduction is 4.5%, 4.4% and 3.0%.

Another important observation is the fact that the user-time K-NN method using the decay ratings produced significantly lower error values compared to the ratings. The decay ratings reduced the MAE by 24.4% for the complete day, 11.2% for the morning interval and 11.0% for the evening interval.

**Table 4.** Values of the RMSE, MAE and NMAE obtained for the rating prediction in different time periods using the user-time K-NN method.

| User-time K-NN | MAE | | | RMSE | | | NMAE | |
|---|---|---|---|---|---|---|---|---|
| | Ratings | Decay Ratings | Improv. | Ratings | Decay Ratings | Improv. | Ratings | Decay Ratings |
| Day (24 h) | 1.202 | **0.909** | **24.38%** | 1.506 | **1.121** | **25.56%** | 0.301 | **0.227** |
| Morning | 0.990 | **0.879** | **11.21%** | 1.223 | **1.089** | **10.96%** | 0.247 | **0.220** |
| Evening | 1.031 | **0.918** | **10.96%** | 1.282 | **1.089** | **15.05%** | 0.258 | **0.229** |

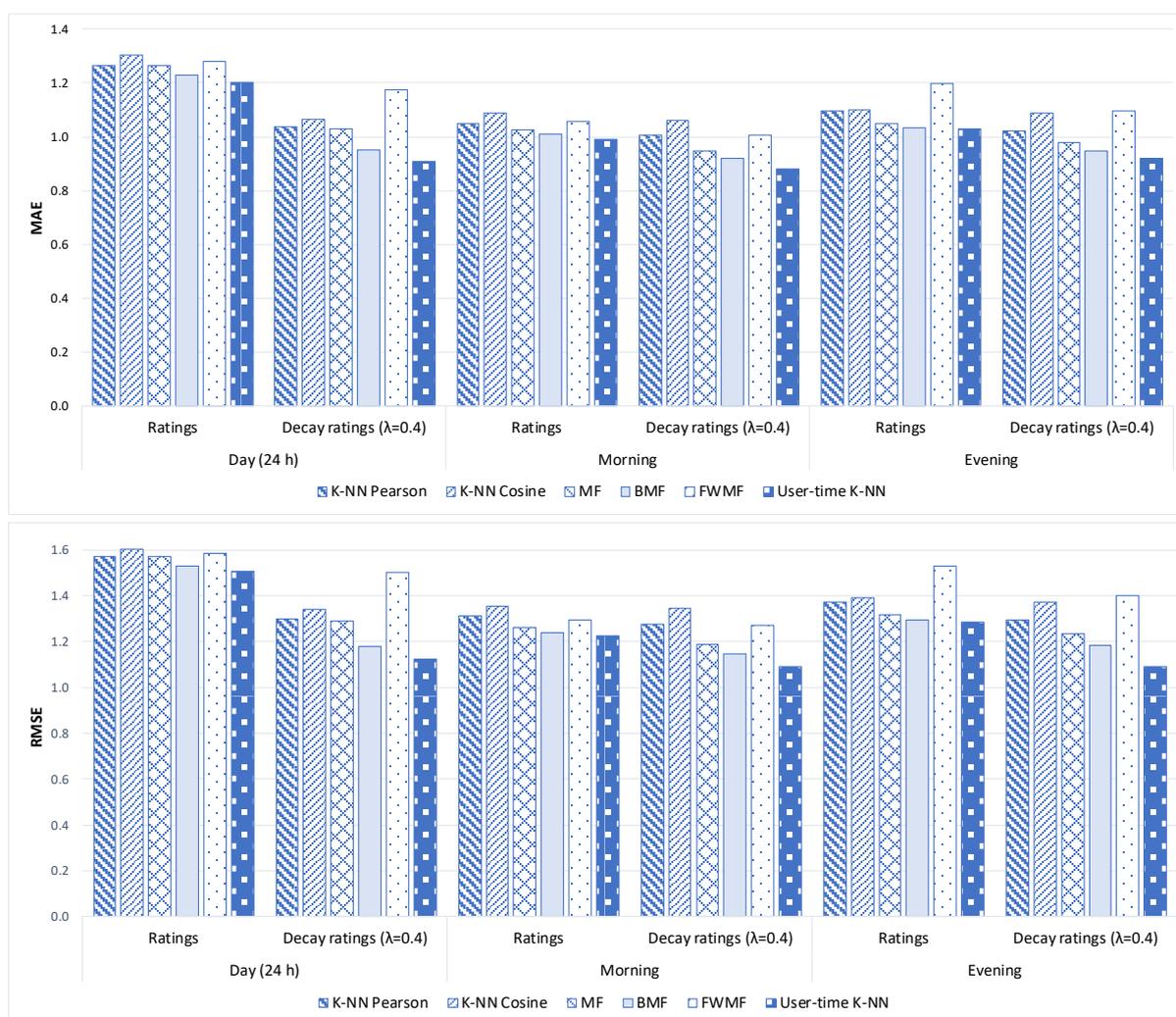

**Figure 7.** The MAE and RMSE values obtained using the time of day as an additional attribute in collaborative filtering (CF (user-time K-NN)) versus the simple use of ratings or decay ratings.

The MAE and RMSE are represented in a different way in Figure 8 in order to analyze the effects of introducing contextual information regarding the time of the day versus the complete day. It can be seen that the error rates were significantly lower when information about the morning and evening periods was introduced, as compared to complete day, when using the ratings. Nevertheless, this reduction was not detected during the evening period using K-NN cosine and user-time K-NN methods and when using the decay ratings. In addition, the reduction generated with the decay ratings for the rest of the cases (all methods for the morning period and K-NN Pearson, MF and BMF for the evening period) was smaller. However, all methods using decay ratings give lower error rates than the same method using ratings. Moreover, our proposal of using decay ratings and the time of day as an additional attribute (user-time K-NN method), was the approach that produced the lowest values of the MAE and RMSE. Specifically, the improvement in the MAE, compared to the BMF with ratings, was 26.0% for the complete day, 11.2% for the morning interval and 12.8% for the evening interval.

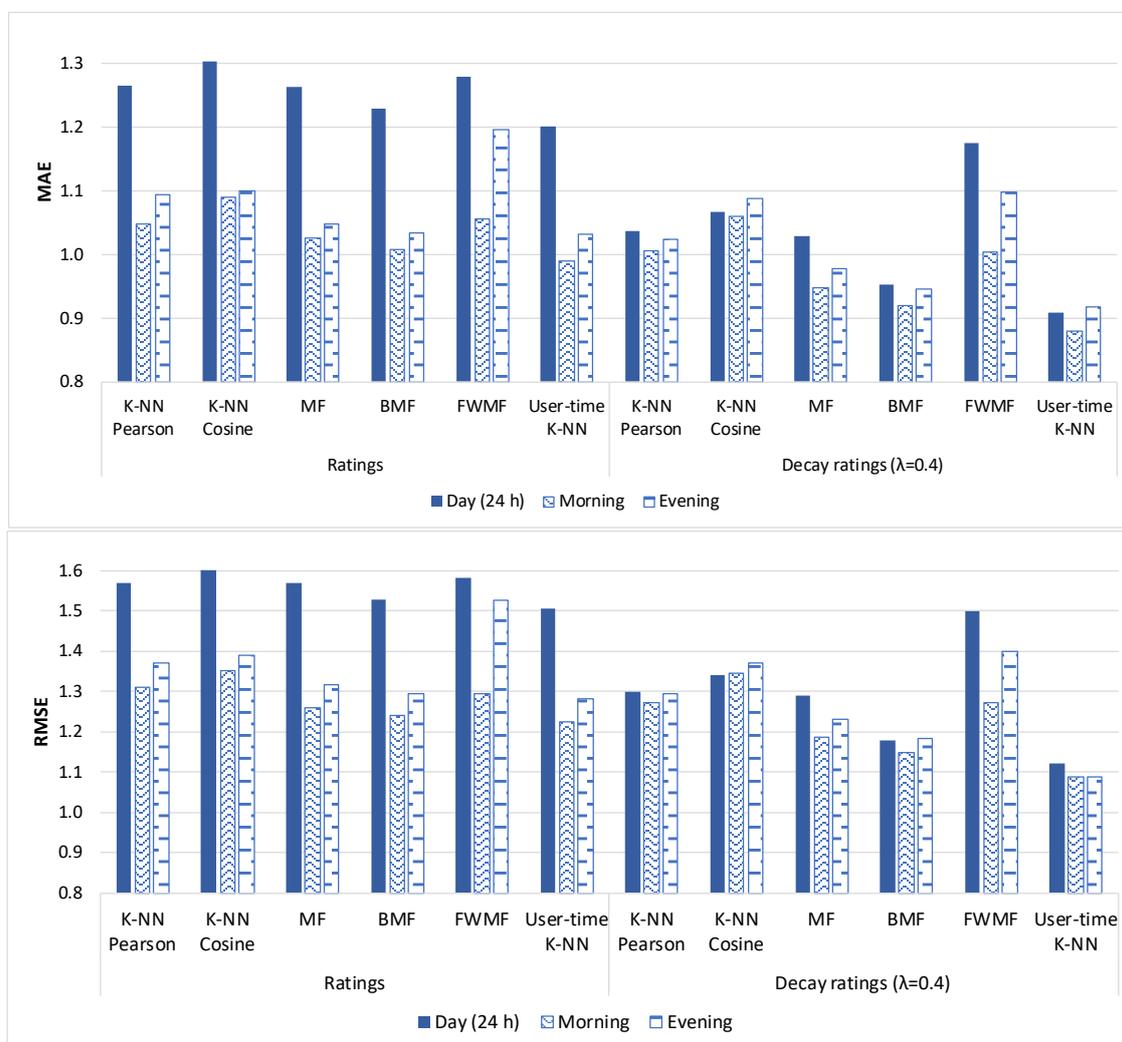

**Figure 8.** Comparison of the MAE and RMSE values obtained for the three periods studied using different methods.

4.4.2. Top-N Recommendations

The effect of introducing the time of day attribute using the user-time K-NN method on top-N recommendations is analyzed in this section. Table 5 shows the AUC, NDCG and MAP obtained by applying that method for the three periods studied, both with the ratings and decay ratings since in both cases the results were the same. Figure 9 shows the values of these evaluation metrics, as well as prec@5, prec@10, prec@15 and MAP, obtained after applying the proposed user-time K-NN method and other top-N recommendation approaches, with both the ratings and decay ratings and considering the context of time of day.

In the same way as applying the K-NN method without attributes, applying the user-time K-NN method produced the same results of the rank-based metrics for the ratings and decays ratings. This occurs even though the MAE and RMSE error rates were lower with the ratings than with the decay ratings. This was possible because the MAE and RMSE are based on the difference between predicted values and actual values, while the metrics used for top-N recommendations are based on the items included in the ranked lists and their ranks. After analyzing the top-N lists in detail we found few changes in their position, especially in the first elements. The NDCG is a metric not very sensitive to changes that do not affect the first items on the list. These small differences also mean that the proportion of relevant items in the lists, as well as in the top-5, top-10 and top-15 positions, remain almost unchanged. As a result, the differences in the values of all rank-based metrics for ratings and decay ratings are in non-significant digits, even though the overall error rate was lower for the decay ratings.

**Table 5.** The AUC, NDCG and MAP values obtained for top-N recommendation considering the context of time of day.

| User-time K-NN | AUC | NDCG | MAP |
|---|---|---|---|
| | Ratings/Decay Ratings | Ratings/Decay Rating | Ratings/Decay Rating |
| Day (24 h) | 0.708 | 0.348 | 0.009 |
| Morning | 0.662 | 0.331 | 0.008 |
| Evening | 0.692 | 0.551 | 0.001 |

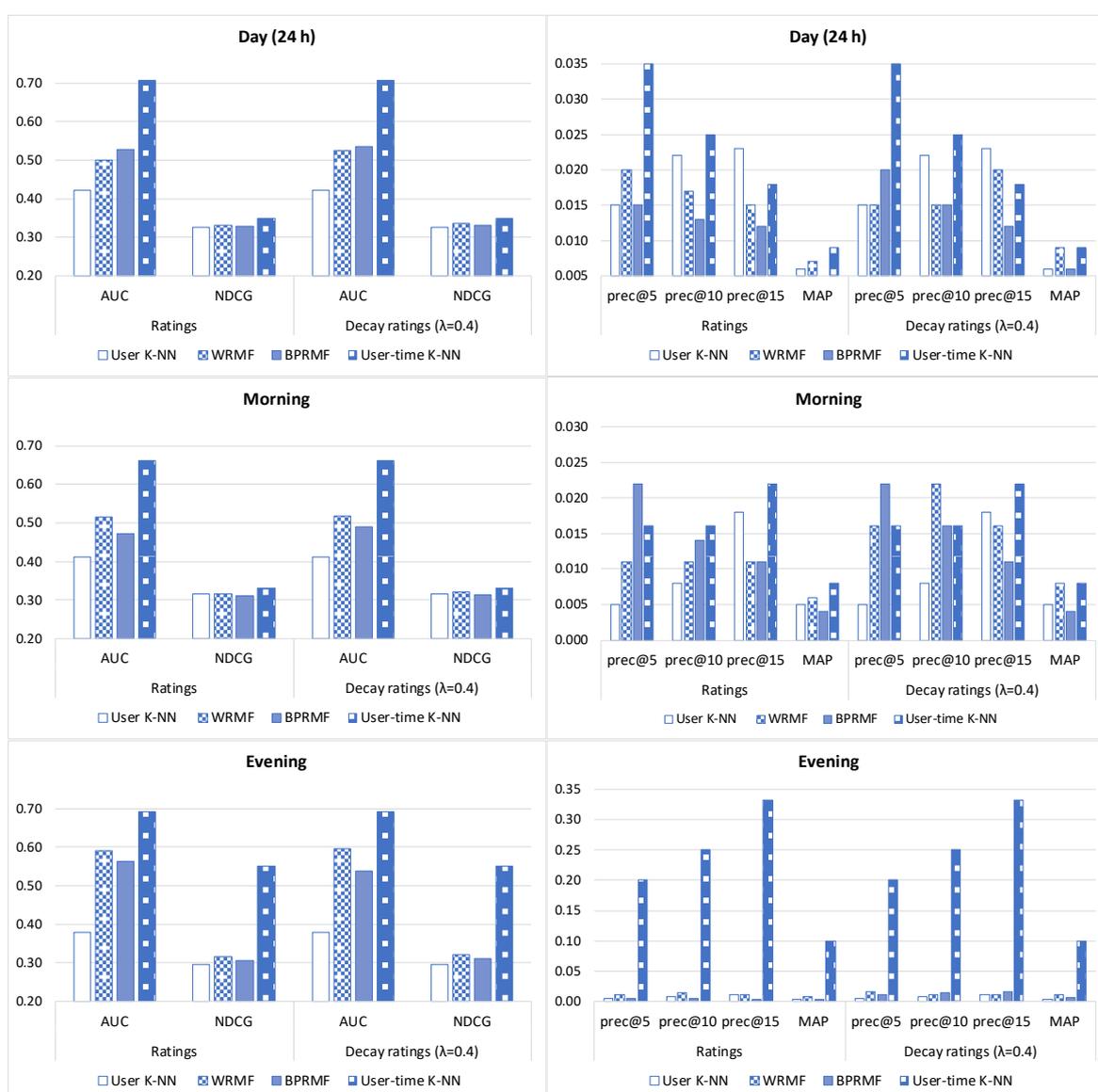

**Figure 9.** Comparison of the top-N metrics obtained using the proposed user-time K-NN method versus the other recommendation methods for the different time periods studied. Comparison using ratings and decay ratings.

Figure 9 shows that most of the metrics were significantly improved by introducing the time of day attribute when using the proposed user-time K-NN method for both the complete day and the morning and evening intervals. The AUC and NDCG values are significantly higher in all cases for both the ratings and decay ratings. The metrics, prec@5, prec@10, prec@15 and MAP, also present much higher values during the evening intervals, which is not always the case for the complete day and the morning intervals. The results are therefore not as good for the top-N lists as they are for the rating prediction. This can also be explained by the reasons given before, which are related to the effect on the rank-based metrics of the closeness of the implicit evaluation values for the relevant songs, as well as the large number of songs contained within the dataset.

## 5. Conclusions

One of the most influential factors in music recommendation is time. Moreover, its influence can depend on various aspects such as the evolution of user preferences and the time when users listen to songs. These aspects have been studied in-depth, but as far as we know, the evolution of preferences has never been modeled directly from implicit feedback. In addition, time has also been considered as a contextual variable in other studies in order to make recommendations based on the time of day, but it is not used to characterize users according to their listening habits.

In this paper, an approach is proposed to address the effects of time on music recommendations. The method is designed to deal with problems characteristic of the music field which are absent in other application domains. The study also attempts to manage implicit feedback, since explicit ratings are not usually available. Thus, the evolution of user preferences over time is included in the proposed method for obtaining the implicit ratings of songs from the frequency of plays. Other commonly used methods for incorporating time dynamics, such as decay functions, cannot be used here because they require timestamped explicit ratings. Likewise, the traditional way of computing implicit ratings, such as using the time users spend on examining an item or the purchase information, are not applicable in the music domain. This occurs because users tend to listen to a song more than once and the songs are not bought one by one on the streaming platforms. Furthermore, the proposed recommender model incorporates information from the listening behavior of users in order to improve recommendations.

The results of the validation performed using a dataset from Last.fm proved that the proposed method for recommending songs outperforms other collaborative filtering approaches, when the context of time is or is not taken into account.

**Author Contributions:** Conceptualization, M.N.M.-G. and Y.Z.; methodology, D.S.-M., M.N.M.-G. and Y.Z.; software, D.S.-M.; validation, D.S.-M. and M.N.M.-G.; data curation, D.S.-M.; supervision, Y.Z. and M.N.M.-G.; funding acquisition, M.N.M.-G. All authors have read and agreed to the published version of the manuscript.

**Funding:** This research was funded by the Junta de Castilla y León, Spain, grant number: SA064G19.

**Conflicts of Interest**: We declare no conflict of interest